\documentclass[twocolumn,amsmath,amssymb,prl,superscriptaddress,longbibliography]{revtex4-1}

\usepackage{titlesec}
\usepackage[colorlinks,bookmarks=false,citecolor=blue,linkcolor=red,urlcolor=blue]{hyperref}
\usepackage{epsfig}
\usepackage{color}
\usepackage{graphicx} 
\usepackage{dcolumn}  
 \usepackage{amsmath}
\usepackage{epstopdf}

\usepackage{caption}

\usepackage[utf8]{inputenc}

\newcommand{\be}{\begin{equation}}
\newcommand{\ee}{\end{equation}}

\definecolor{drkgr}{rgb}{0.05,0.6,0.2}

\begin{document}

\title{Energy scales in 4$f^1$ delafossite magnets:
       crystal-field splittings larger than the strength of spin-orbit coupling in KCeO$_2$}

\author{M.~S.~Eldeeb}
\affiliation{Institute for Theoretical Solid State Physics, Leibniz IFW Dresden, Helmholtzstr.~20, 01069 Dresden, Germany}

\author{T.~Petersen}
\affiliation{Institute for Theoretical Solid State Physics, Leibniz IFW Dresden, Helmholtzstr.~20, 01069 Dresden, Germany}

\author{L.~Hozoi}
\affiliation{Institute for Theoretical Solid State Physics, Leibniz IFW Dresden, Helmholtzstr.~20, 01069 Dresden, Germany}

\author{V.~Yushankhai}
\affiliation{Joint Institute for Nuclear Research, Joliot-Curie 6, 141980 Dubna, Russia}
\affiliation{Max-Planck-Institut f\"ur Physik komplexer Systeme, N\"othnitzerstr.~38, 01187 Dresden, Germany}


\author{U.~K.~Rossler}
\affiliation{Institute for Theoretical Solid State Physics, Leibniz IFW Dresden, Helmholtzstr.~20, 01069 Dresden, Germany}

\begin{abstract}
\noindent
Ytterbium-based delafossites with effective $\tilde{S}\!=\!1/2$ moments are intensively investigated as
candidates for quantum spin-liquid ground states.
While the synthesis of related cerium compounds has also been reported, many important details concerning
their crystal, electronic, and magnetic structures are unclear.
%
Here we analyze the $\tilde{S}\!=\!1/2$ system KCeO$_2$, combining complementary theoretical methods.
The lattice geometry was optimized and the band structure investigated using density functional theory
extended to the level of a GGA+$U$ calculation in order to reproduce the correct insulating behavior.
The Ce 4$f^1$ states were then analysed in more detail with the help of {\it ab initio} wave-function-based
computations.
Unusually large effective crystal-field splittings of up to 320 meV are predicted, which puts KCeO$_2$
in the strong field coupling regime.
%
%
%
Our results reveal a subtle interplay between ligand-cage electrostatics and the trigonal field generated
by the extended crystalline surroundings, relevant in the context of recent studies on tunning the nature
of the ground-state wave-function in 4$f$ triangular-lattice and pyrochlore compounds.
It also makes KCeO$_2$ an interesting model system in relation to the effect of large crystal-field splittings
on the anisotropy of intersite exchange in spin-orbit coupled quantum magnets.
\end{abstract}

\date\today
\maketitle

{\it Introduction.}
Along with the on-site Coulomb repulsion, spin-orbit coupling is considered to define a dominant energy
scale in $f$-electron compounds.
In 4$f^{13}$ ytterbium oxides and chalcogenides, for example, materials that are investigated intensively
nowadays as candidates for spin-liquid ground states \cite{yb114_li_15_1,yb114_li_15_2,yb112_liu_2018,
yb112_baenitz_2018,yb112_baenitz_2019,yb112_tsirlin_2019,yb112_bordelon_2019,yb112_dai_2020}, the separation
between the low-lying states of the split $J\!=\!7/2$ ground-state multiplet and those of the excited
$J\!=\!5/2$ term is in the range of 1.3 eV.
Comparatively, the splittings induced by the ionic solid-state surroundings imply a scale of tens of
meV \cite{yb112_liu_2018,yb112_baenitz_2018,yb112_baenitz_2019,yb112_tsirlin_2019,yb112_bordelon_2019,
yb112_dai_2020,yb13_ziba_2019}.
A notable feature, however, is that for lighter ligands in these systems, i.\,e., O instead of S or Se,
the crystal-field splittings may increase up to $\approx$100 meV \cite{yb112_baenitz_2019,yb112_tsirlin_2019,
yb112_bordelon_2019,yb13_ziba_2019}. 
This can be qualitatively explained by having shorter M-O bonds, which leads to stronger ligand-field
effects, and also by more subtle chemical aspects giving rise to stronger trigonal compression of the
oxygen cage in the oxides.
Starting from such observations on the Yb-based compounds, in particular, the triangular-lattice
NaYbL$_2$ delafossites \cite{yb112_liu_2018,yb112_baenitz_2018,yb112_baenitz_2019,yb112_tsirlin_2019,
yb112_bordelon_2019,yb112_dai_2020,yb13_ziba_2019}, the Ce-based analogues, e.\,g., KCeO$_2$
\cite{KCeO2_clos_1970}, look from an electronic-structure point of view somewhat more peculiar:
the Ce 4$f$ states are known to be more extended, i.\,e., likely more sensitive to the ligand environment;
on the other hand, the spin-orbit coupling is significantly weaker for early rare-earth ions, by factors
of $\sim$5 for Ce$^{3+}$ as compared to Yb$^{3+}$ \cite{4f_atanasov_16}.
An interesting regime can then be realized where spin-orbit interactions and crystal-field splittings
have similar magnitude.
Situations of this kind were discussed in the context of strong deviations from the $j_{\mathrm{eff}}\!=\!1/2$
picture in $t_{2g}^5$ 5$d$ and 4$d$ quantum magnets such as CaIrO$_3$ \cite{CaIrO3_niko_2012,CaIrO3_marco_2014}
and $\alpha$-RuCl$_3$ under high pressure \cite{RuCl3_gael_2018}, where the trigonal splittings imply a
larger energy scale as compared to the strength of the 5$d$/4$d$-shell spin-orbit coupling;
they modify the nature of the ground-state wave-functions and therefore the relevant intersite exchange
paths.
Looking for related physics in the case of 4$f$ materials, we address in this study the on-site electronic
structure of Ce ions in KCeO$_2$.
Impetus is also provided by recently finding crystal-field splittings of up to 125 meV in the sister
compound KCeS$_2$ \cite{KCeS2_gael_2020}.

{\it Lattice geometry and electronic structure from density functional theory.\,}
While the synthesis of KCeO$_2$ was already reported decades ago \cite{KCeO2_clos_1970}, a complete
characterization of its crystal structure is still missing.
We therefore start our discussion with an analysis of structural aspects in KCeO$_2$ --- for clarifying 
those we rely on density-functional calculations with periodic boundary conditions.
KCeO$_2$ crystallizes in the NaFeO$_2$-type delafossite structure with a rhombohedral lattice (space
group $R\bar{3}m$, no 166).
In hexagonal setting K, Ce, and O have the Wyckoff positions 3$a$ (0,\,0,\,0), 3$b$ (0,\,0,\,1/2), and
6$c$ (0,\,0,\,$z$), respectively (see Fig.\;1).
The experimental room-temperature lattice constants are $a=3.66$ and $c=18.66$~\AA \ \cite{KCeO2_clos_1970}.
The position of O in the cell, i.\,e., the $z$ parameter, has not yet been established.
Using the delafossite setting, we determined this parameter and also performed a complete lattice
optimization.
The full-potential local-orbital code \textsc{fplo} (version 18) \cite{FPLO} was employed for this 
purpose.
As exchange-correlation functional we applied the generalized-gradient approximation (GGA)~\cite{PBE96}.
In the context of lattice optimization for 4$f$ compounds, details on the performance of \textsc{fplo}
and of different functionals were recently published by Majumder {\it et al.} \cite{yb1114_majumder_20}.

We first optimized the $z$ parameter using the experimentally derived lattice constants and three 
different approaches: plain GGA with and without spin polarization and also a GGA+$U$ spin-split
calculation.
For the latter, the Coulomb repulsion parameter was set to $U\!=\!5.0$~eV and the Hund exchange to
$J_H\!=\!0.69$ eV by fixing the Slater parameters for the 4$f$ states of Ce to $F_0\!=\!U$, $F_2\!=\!8.54$,
$F_4\!=\!5.37$, and $F_6\!=\!3.86$~eV. 
The $F_2$ to $F_6$ ratio was adopted from Hartree-Fock calculations for free ions \cite{mann1967atomic};
the value of $J_H$ was renormalized by a factor of 0.7. 
The so-called atomic limit was used as double-counting correction.
A $k$-mesh of 24$\times$24$\times$24 points, corresponding to 13824 irreducible $k$ points, was found to
be sufficiently accurate.

\begin{figure}[b]
\includegraphics[width=0.75\columnwidth]{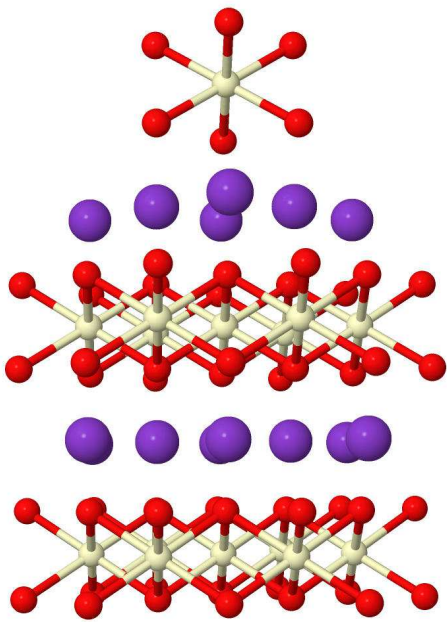}
\caption{
Succesive ionic layers in KCeO$_2$.
Light yellow, red, and purple spheres represent Ce, O, and K sites, respectively.
For the top CeO$_2$ layer, only one single CeO$_6$ octahedron is displayed.
}
\label{fig_struct}
\end{figure}

\begin{figure}[t]
\includegraphics[width=1.4\columnwidth]{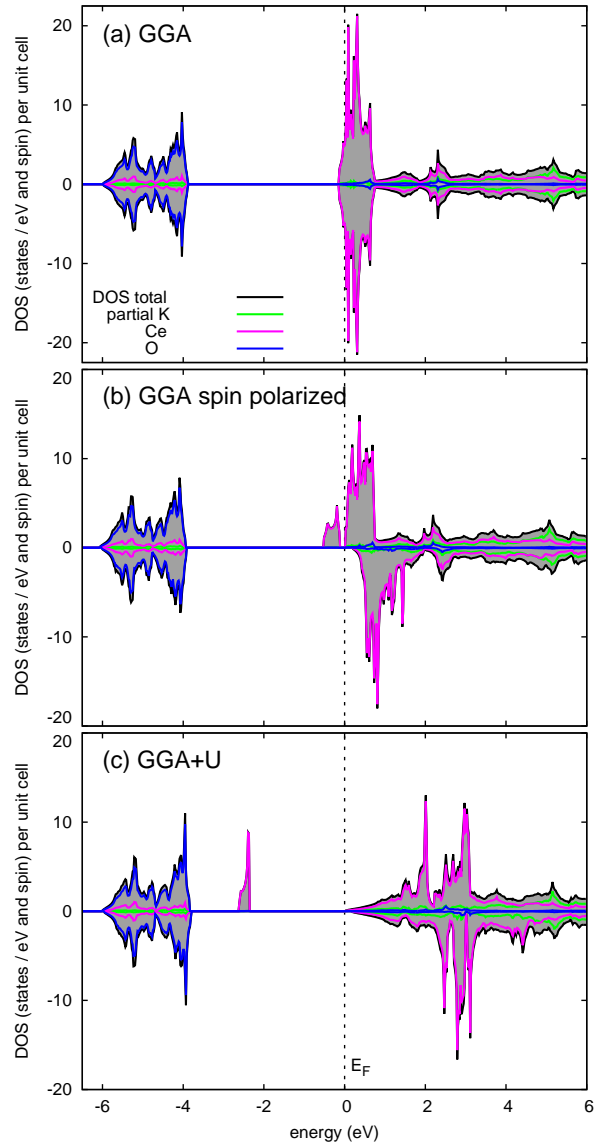}
\caption{
Densities of states (DOS's) in KCeO$_2$ from DFT, including site projected DOS's for the
three different ionic species:
results for spin-degenerate (a) and spin-polarized (b) plain GGA along with (c) GGA+$U$ data.
Spin-majority and minority DOS's are plotted in positive and negative ordinate direction, 
the Fermi level $E_F$ is at zero energy.
}
\label{fig_DOS}
\end{figure}

Results based on density functional theory (DFT) are listed in Table\;\ref{dft_str}.
The data show that accounting for spin polarization has only marginal influence on the lattice geometry
and also that the GGA+$U$ scheme modifies the $z$ parameter by only a small amount.
Plain GGA yields a metallic state as Ce $4f$ bands show up at the Fermi level.
On the other hand, in subsequent GGA+$U$ calculations a finite gap arises. 
The density of states for the three different approaches are compared in Fig.~\ref{fig_DOS}.

The optimization of the lattice parameters at the GGA+$U$ level yields a too large lattice volume, by
about 3.5\% as compared to experimental estimates \cite{KCeO2_clos_1970}.
This matches the usual tendency of the GGA functional to overestimate both volume and spin polarization.
The cohesive energy of the compound is largely determined by the ionic contribution.
The electronic states of Ce, in particular, the 5$d$'s, and their hybridization with O states do
play a certain role and an improved accuracy probably requires improved description of electronic
correlations.
Yet, the relatively small effect on the O-ion position in the different DFT schemes suggests that the
geometry optimization is reasonably robust and can be reliably used for a more detailed analysis of the
Ce 4$f$ electronic structure. 
The value computed by DFT for the O-ion $z$-axis fractional coordinate implies rather strong trigonal
compression of the ligand cages, with O-Ce-O angles deviating by 7.7$^{\circ}$ from 90$^{\circ}$
bond angles.
The impact of this trigonal distortion on the 4$f$-shell energy level structure is discussed in the
next section.
For the quantum chemical calculations, we adopted the O $z$-axis parameter derived by GGA+$U$ when
using the experimental lattice constants.


\begin{table}[b]
\caption{
\label{dft_str}
Structural data obtained for KCeO$_2$ by various types of DFT computations.
For the first four entries the experiental lattice constants were used.
%
Both scalar relativistic and fully relativistic calculations with spin-orbit coupling (SOC) were
performed.
}
\begin{tabular}{l | c c c }
\hline
\hline\\[-0.30cm]
Method                            &$a$       &$c$       &$z$    \\
 \                                &$($\AA$)$ &$($\AA$)$ & \     \\
\hline
\\[-0.20cm]
$\textrm{GGA}$                    &$ $       &$ $       &0.2301 \\
$\textrm{GGA spin polarized}$     &3.66$^a$  &18.66$^a$ &0.2304 \\
$\textrm{GGA+$U$ spin polarized}$ &$ $       &$ $       &0.2310 \\
$\textrm{GGA+$U$+SOC}$            &$ $       &$ $       &0.2311 \\
\\[-0.35cm]
\hline
\\[-0.15cm]
$\textrm{GGA+$U$ spin polarized}$ &3.694     &18.964    &0.2299 \\
$\textrm{GGA+$U$+SOC }$           &3.696     &18.954    &0.2299 \\
\hline
\hline
\end{tabular}

{\ }$^a$ Fixed.
\end{table}

{\it Quantum chemical calculations for the Ce$^{3+}$ 4$f^1$ multiplet structure.}
In the intermediate coupling regime with equally strong spin-orbit and crystal-field interactions, for
describing the energy levels and the corresponding eigenstates of the 4$f^1$ configuration of Ce$^{3+}$,
the full basis of 14 atomic-like spin orbitals should be used.
Besides the value of the spin-orbit coupling $\lambda$, for $D_{3d}$ point-group symmetry
\cite{KCeO2_clos_1970}, six crystal-field parameters are required in the relevant effective model
\cite{wybourne_1965}.
Fitting those effective interaction constants require rather detailed experimental data --- 
$f$-$f$ excitation energies, $g$ factors etc.
The situation becomes delicate when not all $f$-$f$ transitions are captured in, e.g., the neutron
scattering spectra \cite{yb112_baenitz_2018} or additional peaks arise due to vacancies, interstitials,
or strong electron-phonon couplings \cite{KCeS2_gael_2020,Ce227_gaudet_2019}.

The associated uncertainties, however, can be overcome with the help of {\it ab initio} computations
of the 4$f$ multiplet structure.
Using crystallographic data as on the third position in Table\;\ref{dft_str}, we performed
such calculations for a finite set of atoms having a CeO$_6$ unit as central region, in particular,
multiconfiguration and multireference quantum-chemical computations \cite{Helgaker2000} both with and
without spin-orbit coupling, and the main results are reported in Table\;\ref{4f1_dft_str}.
An active space defined by the seven Ce 4$f$ orbitals was employed to this end for the initial
multiconfiguration calculation.
The latter was carried out as a complete-active-space self-consistent-field (CASSCF) optimization
\cite{Helgaker2000} for an average of the seven 4$f^1$ $S\!=\!1/2$ states.
Multireference configuration-interaction (MRCI) wave-functions were subsequently built by additionally
considering single and double excitations \cite{Knowles92} out of the Ce 4$f$ and O 2$p$ orbitals of
the `central' CeO$_6$ octahedron.
Effective core potentials and valence basis sets as optimized in Refs.\;\cite{ECPs_RE_1_dolg_89,BSs_RE_dolg_02}
were used for the central Ce ion, along with all-electron [4$s$3$p$2$d$] Douglas-Kroll basis sets for the adjacent
ligands \cite{BSs_DK_deJong_01}.
To model the charge distribution in the immediate vicinity, we relied on large-core pseudopotentials
including the 4$f$ electrons in the core as concerns the six Ce nearest neighbors \cite{ECPs_RE_2_dolg_89,
BSs_RE_dolg_93} and on total-ion potentials as concerns the twelve adjacent K sites \cite{ECPs_fuentealba_82}.
The remaining part of the extended crystalline surroundings was modeled as an effective electrostatic
field \cite{PCembedd_klintenberg_00}.
To determine the symmetries of the spin-orbit states, we computed the $g$ factors for each of those and
additionally the dipole transition matrix elements for 4$f$$\rightarrow$5$d$ excitations.
For instance, the 4$f^1$ spin-orbit states of $\Gamma_4$+$\Gamma_5$ symmetry can be quickly identified
as those which have $z$-component dipole matrix elements [where $z$ ($c$) coincides with the trigonal axis]
with only two of the 5$d^1$ spin-orbit states \cite{f_orb_kebaili_2012}.
Larger active orbital spaces including both shells, 4$f$ and 5$d$, were used for obtaining the dipole
transition matrix elements.
The $g$ factors were obtained according to the procedure described in Ref.\,\cite{Ir214_niko_15}; by
symmetry, the $g_{ab}$ components vanish for the $\Gamma_4$+$\Gamma_5$ spin-orbit states.
The quantum chemical package {\sc molpro} \cite{molpro12} was employed for all wave-function-based
computations.

\begin{table}[t]
\caption{
Ce$^{3+}$ 4$f^1$ multiplet structure (relative energies in meV) using fractional coordinates as
optimized by DFT (see text), CASSCF and MRCI data without spin-orbit coupling along with spin-orbit MRCI
results (MRCI+SOC).
For the double group, notations as in Ref.\;\cite{sugano_tanabe_70} (e.\,g., Appendix\;I in
\cite{sugano_tanabe_70}) are used.
The ground-state $g$ factors are $g_c\!=\!0.31$ and $g_{ab}\!=\!0.09$.
}
\begin{tabular}{l c c | c l}
\hline
\hline

              &CASSCF &MRCI  &MRCI+SOC &\\

\hline

$^2\!A_{2u}$  &0      &0     &0        &$\Gamma_6$\\
$^2\!E_{u}$   &91     &96    &121      &$\Gamma_4$+$\Gamma_5$\\
              &91     &96    &143      &$\Gamma_6$\\
$^2\!A_{1u}$  &132    &131   &252      &$\Gamma_6$\\
$^2\!E_{u}'$  &226    &229   &352      &$\Gamma_4$+$\Gamma_5$\\
              &226    &229   &395      &$\Gamma_6$\\
$^2\!A_{2u}'$ &314    &318   &469      &$\Gamma_6$\\
\hline
\hline
\end{tabular}
\label{4f1_dft_str}
\end{table}

In the absence of spin-orbit interactions, an octahedral ligand field with full cubic symmetry splits
the $f$ levels into three sets, $a_{2u}$, $t_{2u}$, and $t_{1u}$, each of the latter two being triply
degenerate.
The $a_{2u}$ orbital has its lobes normal to the facets of the ligand octahedral cage and therefore
the lowest energy because the Coulomb repulsion with electronic charge at the ligand sites is minimized;
the $t_{1u}$ orbitals, on the other hand, point directly toward the ligands and are of highest energy
\cite{f_orb_griffith_57}.
%
%
Since the smallest energy scale is defined by the $a_{2u}$--$t_{2u}$ splitting 
\footnote{There are rather detailed studies providing estimates for the $a_{2u}$--$t_{2u}$ splitting
in Ce halides with $O_h$ symmetry, see e.g. Table\;3 in Ref.\;\cite{f_orb_bonding_lukens_2013} and
Fig.\;S1 in Ref.\;\cite{4f_atanasov_16}. It implies a scale of $\sim$30 meV according to {\it ab initio}
data reported in Ref.\;\cite{4f_atanasov_16}.},
the $a_{2u}$ and $t_{2u}$ contributions to the ground-state spin-orbit wave-function do not differ by
much for cubic octahedral environment of the Ce ion \cite{f_orb_bonding_lukens_2013}.
Lowering the 4$f$-site symmetry to trigonal ($D_{3d}$ symmetry in the delafossite structure), the
three-fold degeneracy of the $t_{2u}$ and $t_{1u}$ states is lifted to yield $a_{1u}+e_{u}$ and
$a_{2u}+e_{u}$ sets, respectively.
As concerns the low-lying crystal-field levels in KCeO$_2$, surprisingly large $a_{2u}$--$e_u$ and
$a_{2u}$--$a_{1u}$ splittings of 96 and 131 meV are computed by MRCI (see Table\;\ref{4f1_dft_str}).
This provides a ground-state spin-orbit wave-function that has significantly stronger contribution, 65\%,
from the lowest $^2\!A_{2u}$ trigonal-field term and only $\sim$35\% from $^2\!E_{u}$ and $^2\!A_{1u}$.
Since in this way the in-plane $a_{2u}$--$a_{2u}$ superexchange is enhanced, the result is relevant to the
analysis of the effective magnetic couplings (see \cite{ce01_SE_mironov_96,yb13_SE_palii_03,yb13_rau_2018,kitaev_d_f_motome_2020}
for recent discussion of superexchange paths in rare-earth oxides and chalcogenides).
%

Having so strong crystal-field effects also gives rise to large excitation energies for the low-lying
spin-orbit states --- by MRCI+SOC calculations \cite{SOC_molpro}, we find that the second and third
spin-orbit doublets (relative energies of 121 and 143 meV in Table\;\ref{4f1_dft_str}) are separated
by roughly the same amount from the ground state and from the next excited Kramers doublets.
For a free ion, these next excited states are part of the $^2F_{7/2}$ manifold.
While degenerate in vacuum, the higher-lying eight spin-orbit states cover an energy window of more
than 200 meV in KCeO$_2$.
For completeness, we also performed a calculation for a free Ce$^{3+}$ ion, using the same basis sets
and quantum chemical program.
The $^2F_{5/2}$\,--\,$^2F_{7/2}$ splitting is $\Delta_{\mathrm{SOC}}\!=\!250$ meV by spin-orbit CASSCF.
It implies a spin-orbit coupling constant $\lambda\!=\!2\Delta_{\mathrm{SOC}}/7\!=\!71$ meV
\footnote{Somewhat smaller than estimates for Ce halides \cite{4f_atanasov_16}.}, smaller by $\approx$20
meV than the $a_{2u}$--$e_{u}$ splitting and by a factor of $\approx$4.4 than the $a_{2u}$--$a_{2u}'$
splitting in Table\;\ref{4f1_dft_str}.
This $\lambda$ is actually weaker than the corresponding parameter of, e.\,g., Ru$^{3+}$ 4$d^5$ ions
on the Kitaev honeycomb lattice of $\alpha$-RuCl$_3$ \cite{RuCl3_ravi_2016,RuCl3_markus_2019}.

To put our results in perspective, we note that the lowest two excitation energies (121 and 143 meV,
see Table\;\ref{4f1_dft_str}) are larger by factors of $\gtrsim$4 than the values reported for the Ce
2$p$ halide CeF$_3$ \cite{CeF3_gerlinger_1986}.
While this can be qualitatively understood on the basis of the larger ligand ionic charges in the
oxide, the difference is nevertheless remarkable.
Compared to the sulphide KCeS$_2$ \cite{KCeS2_gael_2020}, the lowest excitation energies are larger 
by factors of 2--3 in KCeO$_2$, matching the trend pointed out by Gerlinger and Schaack when replacing
Cl (3$p$ valence-shell ligand) by F (2$p$ ligand) within the Ce-halide CeX$_3$ family
\cite{CeF3_gerlinger_1986}.

\begin{table}[!t!]
\caption{
Ce$^{3+}$ 4$f^1$ multiplet structure (relative energies in meV) for an idealized cubic CeO$_6$ octahedron,
obtained by appropriately shifting the six ligands along the $z$ coordinate, away from the reference
Ce site.
The ground-state $g$ factors are $g_c\!=\!0.64$ and $g_{ab}\!=\!0.10$.
}
\begin{tabular}{l c c | c l}
\hline
\hline

                &CASSCF &MRCI  &MRCI+SOC &\\

\hline

$^2\!A_{2u}$    &0      &0     &0        &$\Gamma_6$\\
$^2\!A_{1u}$    &37     &38    &117      &$\Gamma_6$\\
$^2\!E_{u}$     &122    &125   &138      &$\Gamma_4$+$\Gamma_5$\\
                &122    &125   &248      &$\Gamma_6$\\
$^2\!E_{u}'$    &212    &215   &363      &$\Gamma_6$\\
                &212    &215   &375      &$\Gamma_4$+$\Gamma_5$\\
$^2\!A_{2u}'$   &220    &225   &425      &$\Gamma_6$\\
\hline
\hline
\end{tabular}
\label{4f1_cubicO6}
\end{table}

Interesting as well is that even for an artificial geometry having the six ligands around the
reference Ce site shifted along the $z$ coordinate away from the Ce ion such that the CeO$_6$
octahedron is cubic, the second and third spin-orbit doublets are still separated from each other
by approximately 20 meV (see Table\;\ref{4f1_cubicO6}).
Since for an isolated cubic octahedron the lowest excited $f^1$ state is a $\Gamma_8$ quartet
(see, e.g., \cite{f_orb_griffith_57,f_orb_bonding_lukens_2013}), this splitting of $\approx$20
meV points to the important role of structural anisotropies beyond the ligand coordination shell,
confirming results of earlier studies on either 4$f$, 5$d$, or 4$d$ compounds \cite{yb13_ziba_2019,
icci_vamshi_2014,Ir214_niko_15,Os227_niko_2013}.
%
The trigonal ligand-cage compression and anisotropies of the extended environment seem
in fact to work in opposite directions as concerns the $e_{u}$--$a_{1u}$ and $e_{u}'$--$a_{2u}'$
splittings: for the cubic octahedron (results in Table\;\ref{4f1_cubicO6}), the $a_{1u}$ level is
lower in energy as compared to the $e_{u}$ states, rather close to the $a_{2u}$ component, while
$a_{1u}$ is above $e_{u}$ with trigonal squeezing of the O$_6$ unit (see Table\;\ref{4f1_dft_str});
a similar trend is seen for the $a_{2u}'$ level, although the latter does not move below $e_{u}'$
when the trigonal squeezing is undone.
The consequence of a reversed sequence of the $e_{u}$ and $a_{1u}$ crystal-field levels is an
inverted sequence of the lowest two spin-orbit excited states (see also the model-Hamiltonian
analysis in Ref.\;\cite{f13_trigonal_perkins_1965}).

Such modulations of the crystal-field splittings, $e_u$--$a_{1u}$ and $a_{2u}$--$a_{1u}$, through
(small) ligand displacements are also of interest in the context of electron-lattice couplings,
i.\,e., the interaction between the nonspherical 4$f$ electronic cloud and optical phonons.
Electron-lattice couplings are known to be strong in Ce compounds (see, e.\,g., the discussion in
Ref.\,\cite{CeF3_gerlinger_1986}).
They were invoked in relation to peculiar features in the Raman spectra of the tysonite trifluoride
CeF$_3$ \cite{CeF3_gerlinger_1986} and are more recently discussed as a possible mechanism behind the
occurence of low-intensity peaks in inelastic neutron scattering experiments on Ce$^{3+}$ pyrochlores
\cite{Ce227_gaudet_2019}.

A related aspect analyzed in Ce$^{3+}$ 4$f^{1}$ pyrochlores is tailoring the $e_{u}$--$a_{2u,1u}$
splittings for realizing a $\Gamma_4$+$\Gamma_5$ on-site spin-orbit ground state, associated with
novel multipolar degrees of freedom and new topological characteristics
\cite{Ce227_chen_2014,Ce227_chen_2017,Ce227_inosov_2020}.
An {\it ab initio} study as performed here on the interplay of ligand-cage distortions and anisotropic
effects involving surroundings beyond the ligand coordination shell is also of interest for pyrochlore
4$f$ compounds since it would better define the conditions under which the $\Gamma_4$+$\Gamma_5$ ground
state can be obtained.
Important structural details in pyrochlore Ce$^{3+}$ systems are (i) having two additional ligands on
the trigonal axis (the ligand cage is defined by eight O ions in pyrochlores) and (ii) having less
pronounced ionic charge imbalance
\footnote{This is a main aspect discussed in, e.g., Refs.\;\cite{yb13_ziba_2019,Os227_niko_2013}.}
between the two different types of cation species in the immediate neighborhood (formally 4+
transition-metal and 3+ Ce nearby sites in the pyrochlores vs 3+ Ce and 1+ alkali nearby cations in
the delaffosite structure).
These structural features in principle destabilize the $a_{2u}$ and $a_{1u}$ orbitals with respect to
the $e_u$ components.
A $\Gamma_4$+$\Gamma_5$ ground state can then be easily envisaged in 4$f^1$ pyrochlores but does not
seem likely in layered triangular-lattice compounds \cite{Ce227_chen_2016}.

We also note that the corrections brought by MRCI to CASSCF are tiny, much less than in the case of
4$f^{13}$ delafossites \cite{yb13_ziba_2019}.
This can be to large extent understood on the basis of the small number of electrons within the $f$
shell; it also indicates that O-to-Ce charge-transfer effects do not play an important role 
\cite{f_orb_bonding_lukens_2013}.
Good agreement is therefore expected with experimental data on the on-site $f$-$f$ excitations,
coming from either inelastic neutron scattering or Raman spectroscopy.
In the context of growing interest in the research area of 4$f$ delafossite-structure quantum magnets,
with extensive literature already available on $\tilde{S}\!=\!1/2$ 4$f^{13}$ delafossites
\cite{yb112_liu_2018,yb112_baenitz_2018,yb112_baenitz_2019,yb112_tsirlin_2019,yb112_bordelon_2019,
yb112_dai_2020}, our analysis provides useful {\it ab initio} benchmarks for the electronic structure
of `complemental' $\tilde{S}\!=\!1/2$ 4$f^{1}$ compounds.

{\it Conclusions.}
In sum, we present an {\it ab initio} investigation of the Ce $f$-shell multiplet structure in the
triangular-lattice compound KCeO$_2$.
Using atomic positions as obtained by DFT lattice optimization, remarkably large crystal-field splittings
are subsequently computed by wave-function-based quantum chemical methods.
A regime that appears unusual for $f$-electron materials is realized this way, in which the 
splittings among the 4$f$ levels as a result of anisotropic surroundings, up to $\approx$320 meV,
are larger than both the spin-orbit coupling constant, $\lambda\!\approx\!70$ meV, and characteristic
free-ion $^2F_{5/2}$\,--\,$^2F_{7/2}$ splitting, $\Delta_{\mathrm{SOC}}\!=\!250$ meV.
%
%
%
%
It remains to be seen how such a setting affects intersite spin interactions, through calculations
based on either effective superexchange models \cite{ce01_SE_mironov_96,yb13_SE_palii_03,yb13_rau_2018,kitaev_d_f_motome_2020}
or on {\it ab initio} methods \cite{Na2IrO3_vmk_14,RuCl3_ravi_2016,SE_chibotaru_15,SE_gendron_19,SE_stoll_19}.
Crystal-field splittings as large as the strength of the spin-orbit coupling are also realized under
high pressure in, e.\,g., the 4$d^5$ Kitaev honeycomb compound $\alpha$-RuCl$_3$;
by modifying the composition of the ground-state wave-function and the dominant exchange paths, they
favor Heisenberg antiferromagnetism in that case \cite{RuCl3_gael_2018}.
Such findings in 4$d^5$ materials suggest that KCeO$_2$ is an interesting model system in the 4$f^1$
category.

 \ 

{\it Acknowledgements.\,}
We thank D.~Inosov, P.~Fulde, and M.~Richter for stimulating discussions, U.~Nitzsche for technical
assistance, and the German Research Foundation (grant HO-4427/3) for financial support.

\bibliography{refs_nov16}

\end{document}